\documentclass[aps,prd,twocolumn,showpacs,10pt,superscriptaddress,preprintnumbers,nofootinbib]{revtex4-2}

\usepackage[utf8]{inputenc}
\usepackage{xspace}
\usepackage[T1]{fontenc}
\usepackage{amsmath,amssymb,amsfonts}
\usepackage{graphicx}
\usepackage{verbatim}
\usepackage[dvipsnames]{xcolor}
\usepackage{nicematrix}
\usepackage{fontawesome5}
\usepackage{ifthen}
\usepackage{lastpage}
\usepackage{subfig}
\usepackage{booktabs}
\usepackage{multirow}
\usepackage{slashed}
\usepackage{wasysym}
\usepackage{orcidlink}
\allowdisplaybreaks

\usepackage{tikz}
\usetikzlibrary{decorations.pathmorphing}
\usetikzlibrary{decorations.markings}
\usetikzlibrary{calc}
\usetikzlibrary{fit, shapes, tikzmark}
\tikzset{
    vector/.style={decorate, decoration={snake}, draw},
	provector/.style={decorate, decoration={snake,amplitude=2.5pt}, draw},
	antivector/.style={decorate, decoration={snake,amplitude=-2.5pt}, draw},
    fermion/.style={draw=black, postaction={decorate},
        decoration={markings,mark=at position .55 with {\arrow[draw=black]{>}}}},
    fermionx/.style={draw=black, postaction={decorate},
        decoration={markings,mark=at position .55 with {\arrow[draw=black]{<}}}},        
    fermionbar/.style={draw=black, postaction={decorate},
        decoration={markings,mark=at position .55 with {\arrow[draw=black]{<}}}},
    fermionnoarrow/.style={draw=black},
    higgs/.style={draw=blue},
    gluon/.style={decorate, draw=black,
        decoration={coil, amplitude=1.5pt, segment length=3pt}},
    photon/.style={decorate, draw=black,
        decoration={coil, aspect=0,amplitude=1.5pt, segment length=4.5pt}},        
    scalar/.style={dashed,draw=black, postaction={decorate}},
    phi/.style={dashed,draw=red, postaction={decorate}},    
    scalarbar/.style={dashed,draw=black, postaction={decorate},
        decoration={markings,mark=at position .55 with {\arrow[draw=black]{<}}}},
    scalarnoarrow/.style={dashed,draw=black},
    electron/.style={draw=black, postaction={decorate},
        decoration={markings,mark=at position .7 with {\arrow[draw=black]{>}}}},
    positron/.style={draw=black, postaction={decorate},
        decoration={markings,mark=at position .35 with {\arrow[draw=black]{<}}}},        
	bigvector/.style={decorate, decoration={snake,amplitude=4pt}, draw},
}

\providecommand{\hypersetup}[1]{}

\providecommand{\pdfbookmark}[3][]{}

\hypersetup{plainpages=false}
\hypersetup{pdfpagemode=UseOutlines}
\hypersetup{bookmarksnumbered=true}
\hypersetup{bookmarksopen=true}
\hypersetup{pdfstartview=FitH}

\usepackage{hyperref}
\allowdisplaybreaks
\hypersetup{
colorlinks=true,
linkcolor=blue,
linktocpage=true,
citecolor=violet,
urlcolor=blue
}
\newcommand{\dd}{\mathrm{d}}

\newcommand{\iu}{\mathrm{i}}

\newcommand{\liverpool}{Department of Mathematical Sciences, University of Liverpool, Liverpool L69 3BX, United Kingdom}

\begin{document}

\title{Tensor decomposition of $e^+e^-\to\pi^+\pi^-\gamma$ to higher orders in the dimensional regulator}
\author{Thomas Dave\,\orcidlink{0009-0003-7880-8229}\,
}
%\email{tcdave@liverpool.ac.uk}
\author{J\'er\'emy Paltrinieri\,\orcidlink{0000-0003-4226-7056}\,
}
%\email{jeremy.paltrinieri@liverpool.ac.uk}
\author{Pau Petit~Ros\`as\,\orcidlink{0009-0009-8824-5208}\,
}
%\email{paupetit@liverpool.ac.uk}
\author{William J. Torres~Bobadilla\,\orcidlink{0000-0001-6797-7607}\,
}
%\email{torres@liverpool.ac.uk}
\affiliation{\liverpool}

\begin{abstract}
We present a first study of the scattering process $e^+ e^-\to\pi^+\pi^-\gamma$ beyond next-to-leading order, aimed at providing preliminary insights required for future NNLO predictions for radiative return processes. A complete four-dimensional tensor decomposition of the amplitude is developed, and the associated one-loop polarised amplitudes are evaluated analytically to higher orders in the dimensional regulator, as required for NNLO accuracy. The calculation is complemented by an efficient numerical framework for the evaluation of the resulting five-point Feynman integrals, enabling stable and fast evaluations across the physical production region with evaluation times of a few hundreds of milliseconds. 
These results are suitable for implementation in Monte Carlo event generators.
\end{abstract}

\maketitle

\section{Introduction}

Precision measurements in electron-positron collisions at low energies provide a cornerstone for testing the Standard Model in regimes where non-perturbative hadronic effects play a dominant role~\cite{Aliberti:2025beg}. 
Theoretical corrections to electron-positron annihilation into hadrons accompanied by an energetic
photon (referred to as radiative return process) are of particular importance, as they enable the extraction of hadronic cross sections at a range of low center-of-mass energies, as has been exploited by experiments such as 
BaBar~\cite{BaBar:2012bdw}, Belle II~\cite{Mori:2007bu}, BESIII~\cite{BESIII:2015equ}, and KLOE~\cite{KLOE:2008fmq, KLOE:2010qei, KLOE:2012anl, KLOE-2:2017fda}.
The precise determination of the hadronic cross section obtained through these measurements enters directly into the evaluation of the hadronic vacuum polarisation contribution to the anomalous magnetic moment of the muon $(g-2)_\mu$~\cite{Muong-2:2025xyk, Muong-2:2023cdq, Muong-2:2021ojo}. The persistent tension between data driven theory and Lattice QCD has elevated the need for improved theoretical predictions to an unprecedented level of accuracy. In this context, reducing theoretical uncertainties in radiative return processes is essential, and requires going beyond the current state-of-the-art in perturbative calculations~\cite{Aliberti:2024fpq}.

While next-to-leading order (NLO) predictions for $e^+e^- \to \pi^+\pi^-\gamma$ are established and implemented in Monte Carlo generators~\cite{Campanario:2019mjh,Budassi:2026lmr,PetitRosas:2026iuq,CarloniCalame:2026hhy}, achieving the precision demanded by current and future measurements calls for next-to-next-to-leading order (NNLO) accuracy. This step is, however, far from trivial. It requires not only the evaluation of genuine two-loop corrections, but also a precise control over the structure of one-loop amplitudes at higher orders in the dimensional regulator.

Recent progress in high-energy physics has demonstrated that one-loop amplitudes expanded to higher orders in the dimensional regulator $\epsilon$ play a crucial role in the construction of NNLO predictions. In particular, the computation of five-point amplitudes through $\mathcal{O}(\epsilon^2)$ has emerged as a key intermediate step toward two-loop results, while simultaneously revealing the analytic structure and functional basis underlying these processes. This program has been successfully carried out for a variety of $2 \to 3$ processes~\cite{Badger:2022mrb,Buccioni:2023okz,Becchetti:2025osw,Bera:2025upg}, establishing a new standard in the precision frontier of collider physics.

In contrast to the high-energy regime, low-energy processes such as radiative return probe a qualitatively different domain. Here, multiple scales and hadronic dynamics interplay in a delicate manner, and theoretical predictions must achieve a level of numerical stability and efficiency compatible with their implementation in Monte Carlo event generators~\cite{PetitRosas:2025xhm}. As a result, the challenges are not only analytic but also computational: {\it fast and reliable evaluation} of five-point amplitudes becomes a central requirement.

In this work, we initiate a systematic study of the evaluation of the scattering amplitude
$e^+ e^-\to\pi^+\pi^-\gamma$ beyond NLO. 
Our primary goal is to provide the building blocks necessary for future NNLO calculations of radiative return processes. At the same time, we aim to develop a framework for the efficient numerical evaluation of the corresponding five-point amplitudes, with a view toward their integration in Monte Carlo event generators.

The presence of multiple kinematic scales in five-point scattering processes, including the full dependence on both lepton and hadron masses, leads to an increase in complexity in the calculation of loop contributions. In this work, we devote significant effort to identify representations that lead to a substantial simplification of the amplitude.
In particular, we employ a tensor decomposition that exploits the fact that all external states live in four space-time dimensions~\cite{Chen:2019wyb,Peraro:2019cjj,Peraro:2020sfm}. While the number of independent tensor structures is  determined by the number of independent helicity configurations of the external particles, their explicit construction requires special care. A naive basis choice typically introduces spurious Gram determinants in the coefficients of the associated form factors, leading to numerical instabilities. To avoid this issue, we construct a basis of tensor structures that is free of such spurious terms, thereby ensuring a stable and efficient representation of the amplitude.
We further elucidate this construction by explicitly working at the level of polarised amplitudes, employing the spinor-helicity formalism~\cite{Kleiss:1985yh}. This approach allows for a compact organisation of the analytic expressions for the form factors, which is particularly well suited for numerical evaluations.

In addition to a suitable representation of the form factors (and polarised amplitudes), an efficient description of the Feynman integrals entering the calculation is essential, especially in view of higher-order contributions. To this end, we adopt the strategy of constructing systems of differential equations for the relevant master integrals that appear in the amplitudes~\cite{Chicherin:2021dyp,Gehrmann:2024tds}, where the dependence on the dimensional regulator is factorised~\cite{Caron-Huot:2014lda}. These equations are then solved numerically~\cite{Boughezal:2007ny,Czakon:2008zk,Mandal:2018cdj,Czakon:2020vql,Czakon:2021yub,Haisch:2024nzv,Badger:2025ljy}, building on the framework developed in~\cite{PetitRosas:2025xhm}.
As a first step in this direction, we focus on one-loop corrections to the radiative return process. We analyse their analytic structure, benchmark their numerical evaluation against existing automated implementations~\cite{Braun:2025afl}, and extend their computation to higher orders in the $\epsilon$-expansion. This provides a first assessment of the functional complexity that is expected to arise in the corresponding two-loop contributions.

This work bridges two directions of current research. On the one hand, it brings modern amplitude techniques into the domain of low-energy precision observables. On the other hand, it emphasises the importance of constructing representations of scattering amplitudes that are not only analytically well understood, but also optimised for numerical evaluations. This interplay between analytic structure and computational efficiency will be essential for achieving NNLO accuracy in low-energy processes.

~

This paper is organised as follows.
In Section~\ref{sec:kin}, we present the analytic structure of the scattering amplitude $e^+ e^- \to \pi^+ \pi^- \gamma$. We discuss its decomposition in terms of tensor structures, together with an efficient construction of polarised amplitudes. We also summarise the ultraviolet (UV) renormalisation and infrared (IR) subtraction required at next-to-leading order.
Section~\ref{sec:1L} outlines the strategy for the computation of the one-loop contribution to the radiative return process. In particular, we elucidate the analytic structure of the relevant Feynman integrals in terms of transcendental functions.
In Section~\ref{sec:num_eval}, we present the numerical evaluation of the polarised amplitudes, and describe an in-house \texttt{C++} implementation that numerically integrates the differential equations satisfied by the transcendental functions.
In Section~\ref{sec:checks}, we present analytic and numerical checks of our results against exiting automated tools. 
Finally, Section~\ref{sec:con} contains our conclusions and an outlook on future directions.
This paper is supplemented by two appendices. Appendix~\ref{app:uv_ir_cts} collects the UV counterterms and IR subtraction constants required for the calculation, expressed in terms of Feynman integrals. Appendix~\ref{app:anc} provides a description of the ancillary files accompanying this work.

\section{Kinematics and tensor decomposition of $\boldsymbol{e^+e^-\to\pi^+\pi^-\gamma}$}
\label{sec:kin}

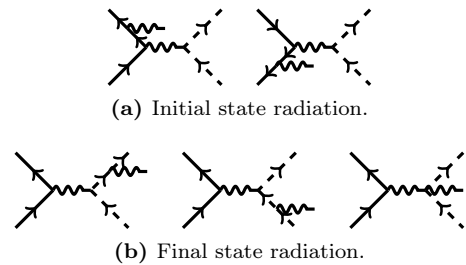
\begin{figure}
\subfloat[\label{fig:isr}Initial state radiation.]{
\begin{tikzpicture}[scale=0.5]
\coordinate (v1) at (0,1);
\coordinate (v2) at (0,-1);
\coordinate (v3) at (3,1);
\coordinate (v4) at (3,-1);
\coordinate (v5) at (1,0);
\coordinate (v6) at (2,0);
\coordinate (a) at (0.5,0.5);
\coordinate (a1) at (1.5,0.5);
\draw[fermion,  very thick] (a) -- (v1);
\draw[fermion,  very thick] (v5) -- (a);
\draw[photon,  very thick] (a) -- (a1);
\draw[fermion,  very thick] (v2) -- (v5);
\draw[photon,  very thick] (v5) -- (v6);
\draw[scalarbar,  very thick] (v3) -- (v6);
\draw[scalarbar,  very thick] (v6) -- (v4);
\end{tikzpicture}
\quad
\begin{tikzpicture}[scale=0.5]
\coordinate (a) at (0.5,-0.5);
\coordinate (a1) at (1.5,-0.5);
\draw[fermion,  very thick] (v1) -- (v5);
\draw[fermion,  very thick] (a) -- (v2);
\draw[fermion,  very thick] (v5) -- (a);
\draw[photon,  very thick] (a) -- (a1);
\draw[photon,  very thick] (v5) -- (v6);
\draw[scalarbar,  very thick] (v3) -- (v6);
\draw[scalarbar,  very thick] (v6) -- (v4);
\end{tikzpicture}
}
\\
\subfloat[\label{fig:fsr}Final state radiation.]{
\begin{tikzpicture}[scale=0.5]
\coordinate (a) at (2.5,0.5);
\coordinate (a1) at (3.5,0.5);
\draw[fermion,  very thick] (v5) -- (v1);
\draw[fermion,  very thick] (v2) -- (v5);
\draw[photon,  very thick] (v5) -- (v6);
\draw[scalarbar,  very thick] (v6) -- (v4);
\draw[scalarbar,  very thick] (v3) -- (a);
\draw[photon,  very thick] (a) -- (a1);
\draw[scalarbar,  very thick] (a) -- (v6);
\end{tikzpicture}
\quad
\begin{tikzpicture}[scale=0.5]
\coordinate (a) at (2.5,-0.5);
\coordinate (a1) at (3.5,-0.5);
\draw[fermion,  very thick] (v5) -- (v1);
\draw[photon,  very thick] (a) -- (a1);
\draw[fermion,  very thick] (v2) -- (v5);
\draw[photon,  very thick] (v5) -- (v6);
\draw[scalarbar,  very thick] (v3) -- (v6);
\draw[scalarbar,  very thick] (v6) -- (a);
\draw[scalarbar,  very thick] (a) -- (v4);
\end{tikzpicture}
\quad
\begin{tikzpicture}[scale=0.5]
\coordinate (a1) at (3,0);
\draw[fermion,  very thick] (v5) -- (v1);
\draw[photon,  very thick] (v6) -- (a1);
\draw[fermion,  very thick] (v2) -- (v5);
\draw[photon,  very thick] (v5) -- (v6);
\draw[scalarbar,  very thick] (v3) -- (v6);
\draw[scalarbar,  very thick] (v6) -- (v4);
\end{tikzpicture}
}
\caption{Tree-level Feynman diagrams contributing to 
$e^+e^-\to\pi^+\pi^-\gamma$.}
\label{fig:trees}
\end{figure}

We consider the scattering process,
\begin{align}
e^-(p_1) + e^+(p_2) \to \pi^-(-p_3)+\gamma(-p_4)+\pi^+(-p_5)\,,
\label{eq:amplitude}
\end{align}
whose tree-level diagrams are shown in Fig.~\ref{fig:trees}. 
All external momenta are considered incoming and satisfying the momentum conservation $p_1+p_2+p_3+p_4+p_5=0$, 
retaining the dependence on the masses of electrons, $p_1^2=p_2^2=m_e^2$,
and pions, $p_3^2=p_5^2=m_\pi^2$. 
The kinematic variables can be parametrised in terms of five independent invariants, for instance,
\begin{align}
\{s_{12},s_{23},s_{34},s_{45},s_{51} \}\,,
\end{align}
with $s_{ij}=(p_i+p_j)^2$, together with the parity-odd Lorentz invariant,
\begin{align}
\text{tr}_5 &= \text{tr}(\gamma^5 \slashed p_4 \slashed p_3 \slashed p_2 \slashed p_1)\,.
\end{align}
For the evaluation of the scattering process~\eqref{eq:amplitude}, we consider the production region corresponding to the $s_{12}$-scattering channel. This physical region is defined by linear constraints on the kinematic invariants,
\begin{align}
p_1^2 >0\,,&& p_2^2 >0\,, && p_3^2 >0\,, && p_5^2 >0\,,
\notag\\
p_1\cdot p_2>0\,, && p_3\cdot p_5>0\,, && p_3\cdot p_4>0\,, && p_4\cdot p_5>0\,,
\notag\\
p_1\cdot p_3<0\,, && p_1\cdot p_4<0\,, && p_1\cdot p_5<0\,, 
\notag\\
p_2\cdot p_3<0\,, && p_2\cdot p_4<0\,, && p_2\cdot p_5<0\,, 
\label{eq:kin-1}
\end{align}
complemented by the following constraints coming from Gram determinants,
\begin{align}
G(p_i,p_j)<0\,, && G(p_i,p_j,p_k)>0\,, 
\notag\\
\text{tr}_5^2=16\,G(p_1,p_2,p_3,p_4)<0\,, 
\label{eq:kin-2}
\end{align}
with $i,j,k\in\{1,\hdots,5\}$ and $i\ne j\ne k$. In turn, the Gram determinants are defined as,
\begin{align}
G(p_{i_1},\hdots, p_{i_n})
=
\text{det}\left(\begin{array}{ccc}
p_{i_{1}}^{2} & \cdots & p_{i_{1}}\cdot p_{i_{n}}\\
\vdots & \ddots & \vdots\\
p_{i_{n}}\cdot p_{i_{1}} & \cdots & p_{i_{n}}^{2}
\end{array}\right)
\,.
\end{align}

The scattering amplitude~\eqref{eq:amplitude} admits a perturbative expansion in the bare coupling constant, 
\begin{align}
\mathcal{A} 
= \left(4\pi\alpha_0\right)^{3/2}\sum_{L=0}^\infty \left(\frac{\alpha_0}{4\pi}\right)^L\,\mathcal{A}^{(L)}\,.
\label{eq:bare_ampl}
\end{align}
At every loop order, we further decompose the amplitude by considering the origin of the external photon. Explicitly, if the photon is emitted from a fermion line (Fig.~\ref{fig:isr}), we refer to as initial state radiation (ISR). Likewise,  if the photon is emitted from a pion line (Fig.~\ref{fig:fsr}), we refer to as final state radiation (FSR), 
\begin{align}
\mathcal{A}^{(L)} = \mathcal{A}^{(L)}_{\text{ISR}}+\mathcal{A}^{(L)}_{\text{FSR}}\,.
\label{eq:ampl_L}
\end{align}
These amplitudes can be further decomposed according to the interaction structure. 
If we denote any contributions to the electric charge that originates from
fermions as $q_e$ and similarly for those from pions as $q_\pi$, 
tree-level contributions become, 
\begin{subequations}
\begin{align}
 \mathcal{A}^{(0)}_{\text{ISR}} ={}& q_e^2q_\pi\,\mathcal{A}^{(0)}_{\text{ISR};21}\,,
\qquad \mathcal{A}^{(0)}_{\text{FSR}} = q_eq_\pi^2\,\mathcal{A}^{(0)}_{\text{FSR};12}\,.
\end{align}
Additionally, when focusing on purely leptonic or hadronic corrections, the scattering amplitudes in Eq.~\eqref{eq:ampl_L} can be further decomposed into initial-state contributions (ISC) and final-state contributions (FSC), following the convention adopted in~\cite{Aliberti:2024fpq}. Within this decomposition, the one-loop contributions to the radiative-return process take the form:
\begin{align}
\mathcal{A}^{(1)} ={}& \mathcal{A}^{(1)}_{\text{ISC}}+\mathcal{A}^{(1)}_{\text{FSC}}\,,\\
\mathcal{A}_{\text{ISC}}^{(1)}={}&q_{e}^{3}q_{\pi}^{2}\,\mathcal{A}_{\text{ISR};32}^{(1)}
\notag\\
&+q_{e}^{4}q_{\pi}\,(\mathcal{A}_{\text{ISR};41}^{(1)}+N_{F}\,\mathcal{A}_{\text{ISR};41'}^{(1)})
\notag\\
&+q_{e}^{3}q_{\pi}^{2}\,(\mathcal{A}_{\text{FSR};32}^{(1)}+N_{F}\,\mathcal{A}_{\text{FSR};32'}^{(1)})\,,
\\
\mathcal{A}_{\text{FSC}}^{(1)}={}&q_{e}^{2}q_{\pi}^{3}\,\mathcal{A}_{\text{FSR};23}^{(1)}
\notag\\
&+q_{e}q_{\pi}^{4}\,\mathcal{A}_{\text{FSR};14}^{(1)}+q_{e}^{2}q_{\pi}^{3}\,\mathcal{A}_{\text{ISR};23}^{(1)}
\,.
\end{align}
\label{eq:ISR_FSR_deco}
\end{subequations}%
Notice that $\mathcal{A}_{\text{ISR};32}^{(1)}$ and $\mathcal{A}_{\text{FSR};23}^{(1)}$ involve two-virtual-photon exchange diagrams, with emission from ISR and FSR, respectively, and therefore require the evaluation of genuine five-point Feynman integrals. All other contributions involve at most four-point integrals, as shall be discussed in Sec.~\ref{sec:1L}.
 
To approximate the interaction between pions and photons, scalar QED (sQED) is employed together with a factorised ansatz in which a vector form factor (VFF) accounts for non-perturbative effects. In practice, this corresponds to the prescription $F_\pi \times \text{sQED}$.
The main motivation for computing these amplitudes is to establish a framework that remains applicable once more sophisticated implementations of the VFF are available within the same setup for a complete NNLO calculation.\footnote{Notice that the very same organisation of these amplitudes as well as their computation can be used for the evaluation of the radiative return process $e^+e^-\to\mu^+\mu^-\gamma$.}

In this context, ISC are considered first, as they are well understood within perturbation theory and  provide a reliable starting point before addressing the more involved hadronic corrections. 
Moreover, we also consider FSC. These contributions must, however, be treated with care, as they correspond to corrections to a quantity that already includes non-perturbative effects.

\subsection{Tensor decomposition \& Polarised amplitudes}

Since the scattering amplitude~\eqref{eq:amplitude} is made of tensor structures comprised of spinors, gamma matrices, and polarisation vectors, we apply a four-dimensional tensor decomposition by taking advantage of the dimensionality of the external momenta. 
In details, we express the scattering amplitude as, 
\begin{align}
\mathcal{A}^{(L)}=\sum_{i=1}^8 \mathcal{F}_i^{(L)}\, \mathcal{T}_i\,,
\end{align}
with, 
\begin{equation}
\begin{aligned}
\mathcal{T}_{1}= & \overline{v}(p_{2},m_{e})\slashed{p}_{3}u(p_{1},m_{e})(p_{1}\cdot\varepsilon_{4})\,,%\nonumber 
\\
\mathcal{T}_{2}= & \overline{v}(p_{2},m_{e})\slashed{p}_{3}u(p_{1},m_{e})(p_{3}\cdot\varepsilon_{4})\,,%\nonumber 
\\
\mathcal{T}_{3}= & \overline{v}(p_{2},m_{e})\slashed{p}_{5}u(p_{1},m_{e})(p_{1}\cdot\varepsilon_{4})\,,%\nonumber 
\\
\mathcal{T}_{4}= & \overline{v}(p_{2},m_{e})\slashed{p}_{5}u(p_{1},m_{e})(p_{3}\cdot\varepsilon_{4})\,,%\nonumber 
\\
\mathcal{T}_{5}= & \overline{v}(p_{2},m_{e})u(p_{1},m_{e})(p_{1}\cdot\varepsilon_{4})\,,%\nonumber 
\\
\mathcal{T}_{6}= & \overline{v}(p_{2},m_{e})u(p_{1},m_{e})(p_{3}\cdot\varepsilon_{4})\,,%\nonumber 
\\
\mathcal{T}_{7}= & \overline{v}(p_{2},m_{e})\slashed{p}_{3}\slashed{p}_{5}u(p_{1},m_{e})(p_{1}\cdot\varepsilon_{4})\,,%\nonumber 
\\
\mathcal{T}_{8}= & \overline{v}(p_{2},m_{e})\slashed{p}_{3}\slashed{p}_{5}u(p_{1},m_{e})(p_{3}\cdot\varepsilon_{4})\,.
\end{aligned}
\label{eq:tensors}
\end{equation}
To identify the relevant tensor structures, we follow the approach presented in~\cite{Peraro:2020sfm}, exploiting the identity for gamma matrices in four dimensions,
$\gamma^\mu=\sum_{i,j=1}^{4} c_{ij}\,\slashed p_i\, p_j^\mu$,
where the coefficients $c_{ij}$ are not relevant for the present discussion.
It should be noted that, in order to obtain this number of independent tensor structures, the reference momentum of the polarisation vector is chosen to be $p_2$, such that $\varepsilon_4 \cdot p_2 = 0$.

{\it Polarised amplitudes} are systematically constructed from the above tensors by decomposing massive momenta into massless ones and profiting from the spinor-helicity formalism~\cite{Kleiss:1985yh}. 
For simplicity, we construct massless momenta as follows
\begin{align}
\begin{aligned}
p_{i}^{\flat} & =p_{i}-x_{i}\,p_{4}\,,\quad x_{i}=\frac{p_{i}^{2}}{2p_{i}\cdot p_{4}}\quad\text{for }i=1,2,3,5.\,,\\
p_{4}^{\flat} & =p_{4}\,,
\end{aligned}
\end{align}
where $(p_{i}^{\flat})^2=0$, and the momentum of the external photon is chosen as the reference momentum for all massive momenta. The explicit construction of all polarised amplitudes is carried out with the aid of \textsc{s@M}~\cite{Maitre:2007jq} and \textsc{SpinorsExtras}~\cite{Kuczmarski:2014ara}.

Because the form factors exhibit a dependence on Gram determinants, already at tree level, the polarised amplitudes are organised such that these denominators cancel, thereby ensuring numerical stability. To achieve this, the polarised amplitudes are constructed as:
\begin{align}
\mathcal{A}_{h}^{(L)} &= \Phi_{h}\,\tilde{\mathcal{A}}_{h}^{(L)}\,,
\label{eq:pol_ampl}
\end{align}
where $\Phi_{h}$ encodes the phase dependence on the polarisation and helicity states of the external particles, for the eight non-vanishing helicity configurations:
\begin{align}
\begin{aligned}
\vec{h} = 
\{&
+++\,,
--+\,,
-++\,,
+-+\,,
\\
&-+-\,,
+--\,,
++-\,,
---
\}\,,
\end{aligned}
\label{eq:helicities}
\end{align}
with,
\begin{align}
\begin{aligned}
&\Phi_{+++}=\frac{1}{\left\langle p_{1}^{\flat}p_{2}^{\flat}\right\rangle \left[p_{4}^{\flat}p_{2}^{\flat}p_{1}^{\flat}p_{4}^{\flat}\right]}\,,
&  & \Phi_{---}=\mathsf{P}\,\Phi_{+++}\,,
\\
&\Phi_{--+}=\frac{1}{\left[p_{4}^{\flat}p_{1}^{\flat}\right]\left[p_{4}^{\flat}p_{2}^{\flat}\right]}\,,
&  & \Phi_{++-}=\mathsf{P}\,\Phi_{--+}\,,
\\
&\Phi_{-++}=\frac{1}{\left\langle p_{1}^{\flat}p_{2}^{\flat}p_{4}^{\flat}\right]\left[p_{2}^{\flat}p_{4}^{\flat}\right]}\,,
&  & \Phi_{+--}=\mathsf{P}\,\Phi_{-++}\,,
\\
&\Phi_{+-+}=\frac{1}{\left\langle p_{2}^{\flat}p_{1}^{\flat}p_{4}^{\flat}\right]\left[p_{1}^{\flat}p_{4}^{\flat}\right]}\,.
&  & \Phi_{-+-}=\mathsf{P}\,\Phi_{+-+}\,.
\end{aligned}
\end{align}
Here $\mathsf{P}$ denotes the parity operator acting on the spinor products.

The free-phase quantities in~\eqref{eq:pol_ampl} are constructed as:
\begin{align}
\tilde{\mathcal{A}}_{h}^{(L)}&=\sum_{i=1}^{8}\left(\mathbb{H}_{hi}^{(0)}+\text{tr}_5\,\mathbb{H}_{hi}^{(1)}\right)\mathcal{F}_i^{(L)}\,,
\label{eq:FF_to_PolAmpl}
\end{align}
where the matrices $\mathbb{H}$ contain rational functions depending only on the kinematic invariants.

 \subsection{UV renormalisation and IR subtraction}
 \label{subsection:UV-IR}

To handle ultraviolet (UV) renormalisation of the bare form factors and infrared (IR) subtraction for obtaining their finite remainders, we organise the computation such that the cancellation of UV and IR poles occurs directly at the integral level. This is achieved by expressing all UV renormalisation and IR subtraction constants in terms of Feynman integrals. Such organisation, combined with the grading of transcendental functions~\cite{Chicherin:2021dyp,Gehrmann:2024tds}, allows for a systematic cancellation of poles---avoiding the need to explicitly compute them during the reconstruction of analytical expressions through numerical evaluations over finite fields~\cite{vonManteuffel:2014ixa,Peraro:2016wsq,Peraro:2019svx}.

We renormalise the form factors according to, 
\begin{align}
\mathcal{F}_{i} = \left(\frac{\mu^2e^{\gamma_E}}{4\pi}\right)^{-\epsilon}\,
Z_F\, Z_\pi\,Z_\gamma^{\frac{1}{2}}\,
\mathcal{F}_{i;0}\,,
\label{eq:uv_ren}
\end{align}
where $Z_F\,,Z_\pi$ and $Z_\gamma$ are the on-shell wave-function renormalisation constants for leptons, pions and photons respectively. We introduce an additional prefactor to avoid unnecessary $\gamma_\text{E}-\log4\pi$ terms (see Sec.~\ref{sec:1L} for normalisation of Feynman integrals). 
The bare masses are renormalised as $m_F^0 = Z_{m_F}m_F$ and $m_\pi^0 = Z_{m_\pi}m_\pi$, and the bare coupling constant, 
\begin{align}
\alpha^0 = 
\left(\frac{\mu^2e^{\gamma_E}}{4\pi}\right)^{-\epsilon}
Z_{\alpha}\,\alpha(\mu)\,,
\label{eq:a_UVren}
\end{align}
with $N_F$ and $N_S$ heavy active lepton and scalars, respectively.
Explicitly, by plugging Eqs.~\eqref{eq:uv_ren} and~\eqref{eq:a_UVren}
in~\eqref{eq:bare_ampl}, the renormalised one-loop contributions take the form, 
\begin{align}
\mathcal{F}_{i}^{(0)} =&{} \mathcal{F}_{i;0}^{(0)}\,,\\
\mathcal{F}_{i}^{(1)} = &{} \mathcal{F}_{i;0}^{(1)} 
+\left(
\frac{3}{2}\delta Z_\alpha^{(1)}
+\frac{1}{2}\delta Z_\gamma^{(1)}
+\delta Z_F^{(1)}
+\delta Z_\pi^{(1)}
\right)\mathcal{F}_{i;0}^{(0)}
\notag\\
&{}+\delta Z_{m_F}^{(1)}\mathcal{F}_{i;\text{CT}_{F}}^{(0)}
+\delta Z_{m_\pi}^{(1)}\mathcal{F}_{i;\text{CT}_\pi}^{(0)}\,,
\end{align}%
where $\mathcal{F}_{i;\text{CT}_{F}}^{(0)}$ and $\mathcal{F}_{i;\text{CT}_{\pi}}^{(0)}$
are tree-level diagrams that account for the mass correction respectively
for electrons and pions. 
Notice that due to the decomposition of the amplitude~\eqref{eq:ISR_FSR_deco},
one can separately perform UV renormalisation for each gauge invariant set of diagrams by properly accounting
for the electron and muon charges, $q_e$ and $q_\pi$. 
We collect all necessary UV renormalisation constants with the explicit dependence on $q_e$ and $q_\pi$ in Appendix~\ref{app:uv_ir_cts}. 

Once we perform UV renormalisation, we still need to deal with IR poles, which can be predicted by 
means of an IR operator~\cite{Catani:2000ef}, 
\begin{align}
\label{eq:renormalised_FF}
\mathcal{F}_{i}^{(1)} 
= &\mathbf{I}(\epsilon,\mu^2;\{p_i,m_i\})\,\mathcal{F}_{i}^{(0)} 
+\mathcal{F}_{\text{fin};i}^{(1)}\,,
\end{align}
where all $\epsilon$-poles are included in the operator $\mathbf{I}$.
Hence, the ultimate goal is the analytic computation of the contributions of all form factors $\mathcal{F}_{\text{fin};i}$ at finite and higher-order terms in $\epsilon$, explicitly, 
\begin{align}
\mathcal{F}_{\text{fin};i}^{(1)} = 
\mathcal{F}_{i;0}
+\epsilon\,\mathcal{F}_{i;1}
+\epsilon^2\,\mathcal{F}_{i;2}
+\mathcal{O}(\epsilon^3)\,.
\end{align}
We provide in Appendix~\ref{app:uv_ir_cts}, the analytic expression of the operator $\mathbf{I}$
in terms of Feynman integrals to immediately account for cancellation of  $\epsilon$-poles at integral level. 

\section{One-loop amplitudes}
\label{sec:1L}

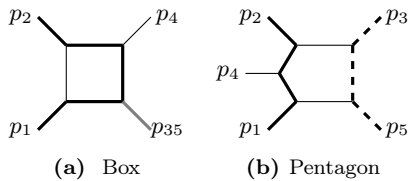
\begin{figure}[t]
\subfloat[\label{fig:box} Box]{
\begin{tikzpicture}[scale=1.5]
\coordinate (v1) at (0.25,0.25);
\coordinate (v2) at (0.25,1.25);
\coordinate (v3) at (1.25,1.25);
\coordinate (v4) at (1.25,0.25);
\coordinate (v5) at (0.5,1);
\coordinate (v6) at (1,1);
\coordinate (v7) at (0.5,0.5);
\coordinate (v8) at (1,0.5);
\draw[fermionnoarrow,  very thick] (v1) -- (v7);
\draw[fermionnoarrow,  very thick] (v2) -- (v5);
\draw[fermionnoarrow,  very thick] (v5) -- (v6);
\draw[fermionnoarrow,  very thick] (v6) -- (v8);
\draw[fermionnoarrow,  very thick] (v8) -- (v7);
\draw[fermionnoarrow,  thin] (v7) -- (v5);
\draw[fermionnoarrow,  thin] (v3) -- (v6);
\draw[draw=gray,  very thick] (v4) -- (v8);
\node  at (0.1,0.25) {\small$p_1$}; 
\node  at (0.1,1.25) {\small$p_2$}; 
\node  at (1.4,1.25) {\small$p_4$}; 
\node  at (1.4,0.25) {\small$p_{35}$}; 
\end{tikzpicture}
}
\subfloat[\label{fig:pentagon}Pentagon]{
\begin{tikzpicture}[scale=1.5]
\coordinate (v9) at (0.35,0.75);
\coordinate (v10) at (0.05,0.75);
\draw[fermionnoarrow,  very thick] (v1) -- (v7);
\draw[fermionnoarrow,  very thick] (v2) -- (v5);
\draw[fermionnoarrow,  thin] (v5) -- (v6);
\draw[fermionnoarrow,  thin] (v8) -- (v7);
\draw[scalar,  very thick] (v6) -- (v8);
\draw[scalar,  very thick] (v3) -- (v6);
\draw[scalar,  very thick] (v4) -- (v8);
\draw[fermionnoarrow,  thin] (v9) -- (v10);
\draw[fermionnoarrow,  very thick] (v9) -- (v5);
\draw[fermionnoarrow,  very thick] (v9) -- (v7);
\node  at (0.1,0.25) {\small$p_1$}; 
\node  at (0.1,1.25) {\small$p_2$}; 
\node  at (1.4,1.25) {\small$p_3$}; 
\node  at (-0.1,0.75) {\small$p_4$}; 
\node  at (1.4,0.25) {\small$p_5$}; 
\end{tikzpicture}
}
\caption{
Independent integral families.
Thick (thin) lines represent massive (massless) momenta, and $p_{35}=p_3+p_5$.
}
\label{fig:int_families}
\end{figure}

We compute the form factors following the standard procedure used in multi-loop calculations.
Tree-level and one-loop Feynman diagrams are generated using 
{\sc FeynArts}~\cite{Hahn:2000kx}
and the corresponding integrands are constructed with {\sc FeynCalc}~\cite{Mertig:1990an,Shtabovenko:2016sxi}.
By applying appropriate projectors from~\eqref{eq:tensors},
we extract the coefficients of the independent tensor structures contributing to the amplitude.
At tree level, we have 5 diagrams (displayed in Fig.~\ref{fig:trees}) while at one-loop
we start with 87 diagrams, however we find that 6 of these diagrams cancel pairwise due to Furry's theorem.
This leaves us with 81 diagrams that contribute to the scattering amplitude at one-loop order.
We find that these diagrams can be systematically decomposed in terms of the independent partial amplitudes~\eqref{eq:ISR_FSR_deco}. We also find that these 81 diagrams can be sorted into 2 integral families (up to crossing of external kinematics) defined by their propagators. 
These families are depicted in Fig.~\ref{fig:int_families} and we provide in Table~\ref{tab:families}  the propagators
that define each family.
We adopt the following normalisation for the Feynman integrals:
\begin{align}
\label{eq:1L_int}
I_{a_1\hdots a_5}^\text{X}=
e^{\gamma_E\epsilon}
\int\frac{d^D\ell}{\imath \pi^{D/2}}\frac{1}{D_1^{a_1}\hdots D_5^{a_5}}\,,
\end{align}
with $\text{X=(P)entagon, (B)ox}$.

\begin{table}[t]
\centering
\begin{tabular}{ccc}
\toprule
Denominator & Box & Pentagon \\
\midrule
$D_1$ &  $(\ell-p_1)^2-m_e^2$ & $\ell^2$ \\
$D_2$ & $\ell^2$ & $(\ell-p_3)^2-m_\pi^2$\\
$D_3$ & $(\ell+p_2)^2-m_e^2$ & $(\ell+p_2)^2-m_e^2$\\
$D_4$ & $(\ell+p_2+p_3)^2-m_e^2$ & $(\ell+p_1+p_2+p_4)^2$\\
$D_5$ & $(\ell+p_2+p_3+p_4)^2$ & $(\ell+p_2+p_4)^2-m_e^2$\\
\midrule
Master Integrals & 8 & 21 \\
\bottomrule
\end{tabular}
\caption{
One-loop independent integral families.
The Feynman prescription $+ \iu0$ is understood for each propagator and not written explicitly.
}
\label{tab:families}
\end{table}

Integration-by-parts identities~\cite{Tkachov:1981wb,Chetyrkin:1981qh,Laporta:2000dsw} are used to reduce the integrals that appear in the form factors
into a minimal set of master integrals that obey a canonical system of differential equations~\cite{Henn:2013pwa}, which were obtained by studying their leading and Landau singularities~\cite{Flieger:2022xyq}. 
We opt to use {\sc LiteRed}~\cite{Lee:2012cn} to generate IBP relations among integrals and master integrals, and then employ {\sc FiniteFlow}~\cite{Peraro:2019svx} to solve over finite fields the resulting linear systems of integrals in terms of master integrals.

Since in this work we are interested in calculating the form factors up to $\mathcal{O}(\epsilon^2)$
we calculate master integrals up to transcendental weight four. 
We note that the master integrals corresponding to the pentagon family shown in Fig.~\ref{fig:pentagon} were recently evaluated in~\cite{PetitRosas:2025xhm}, together with the analytic computation of the interference term, $2\Re(\mathcal{A}^{(1)}_{\text{ISR-TPE};32}\,\mathcal{A}^{(0)*}_{\text{ISR};21})$. 
The remaining master integrals of the box family of Fig.~\ref{fig:box} are computed in this work.

To organise the calculation, we construct a complete system of differential equations for all master integrals contributing to ISC.\footnote{The differential equations for FSC are obtained by crossing of external kinematics.} This includes two pentagon families: one presented in Table~\ref{tab:families}, 
and a second obtained by exchanging $p_3$ and $p_5$. 
In addition, two box families are considered: the one shown in Table~\ref{tab:families} and a second obtained by exchanging $p_1$ and $p_2$. 
Combining these integral families, all contributions can be reduced to a set of 38 master integrals.

Explicitly, the canonical system of differential equations takes the form:
\begin{align}
\dd \overrightarrow{W}_\text{ISC} = \epsilon
\left(\sum_{i=1}^{119}
\mathbb{A}_i\,
 \dd\log\beta_i
\right)
\overrightarrow{W}_\text{ISC}\,,
\label{eq:deq_ISR}
\end{align}
where $\beta$ are the letters of the kinematic alphabet. 
This alphabet is obtained with the aid of {\sc BaikovLetter}~\cite{Jiang:2024eaj} and {\sc Effortless}~\cite{Antonela}, and consists of 119 letters, of which 39 are rational functions and 80 are algebraic. The full alphabet is provided in~\cite{Zenodo}.
The set of integrals $\overrightarrow{W}_\text{ISC}$ already incorporates a rotation of canonical integrals, following the approach of~\cite{Gehrmann:2024tds}, which enables a systematic cancellation of UV and IR poles at the level of Feynman integrals.
To give an indication of the analytic structure of the Feynman integrals, we study the symbol map of the canonical integrals~\cite{Goncharov:2010jf}. In Table~\ref{tab:letters}, we summarise the appearance of letters at different transcendental weights, together with the corresponding number of integrals and transcendental functions.
\begin{table}[t]
\centering
    \begin{tabular}{cc ccc cc cccc cccc cc}
        \toprule
    \multirow{2}{*}{$k$} &&\multicolumn{3}{c}{\# letters}  && \multirow{2}{*}{\# $\overrightarrow{W}_\text{ISC}^{(k)}$} && \multicolumn{4}{c}{$\mathcal{A}_{\text{ISR};32}^{(1)}$} && \multicolumn{4}{c}{$\mathcal{A}_{\text{FSR};23}^{(1)}$}\\
        \cmidrule(lr){2-6}  \cmidrule(lr){8-17}%\cmidrule(lr){11-13}
        &&Even{}  &&Algebraic{}   &&  && $\epsilon^{-1}$& $\epsilon^0$& $\epsilon^1$& $\epsilon^2$ && $\epsilon^{-1}$& $\epsilon^0$& $\epsilon^1$& $\epsilon^2$  \\
        \midrule
		 $0$ &&	-- && -- &&	1 && 0& 0& 0& 0 && 0& 0& 0& 0\\
		 $1$ &&	5 && 6 &&	10&& 4& 4& 0& 0&& 4& 1& 0& 0\\
		 $2$ &&	16 && 17 &&	25&& --& 24& 4& 0&& 0& 24& 1& 0\\
		 $3$ &&	15 && 47 &&	2&& --& --& 26& 4&& 0& 0& 26& 1\\
		 $4$ &&	3 && 0 &&	0&& --& --& --& 26&& 0& 0& 0& 26\\
      \bottomrule
    \end{tabular}
\caption{
Number of letters, canonical integrals contributing to ISC, and transcendental functions
that contribute to $\mathcal{A}_{\text{ISR};32}^{(1)}$ and $\mathcal{A}_{\text{FSR};23}^{(1)}$ 
organised by the weight $k$ at which they first appear in the symbol. 
All letters present at weight $k-1$ also appear at weight 
$k$. No new letters (master integrals) appear beyond weight 5 (3).
}\label{tab:letters}
\end{table}

In details, before inserting the explicit solutions for the integrals, 
the form factors (and polarised amplitudes) become:
\begin{align}
\mathcal{F}^{(1)}_{i} ={}&
\Bigg[
q_eq_\pi \bigg(
\kappa_{15}\,I_{\triangle}(s_{15};m_e^2,m_\pi^2)
+\kappa_{23}\,I_{\triangle}(s_{23};m_e^2,m_\pi^2)
\notag\\
&
-\kappa_{13}\,I_{\triangle}(s_{13};m_e^2,m_\pi^2)
-\kappa_{25}\,I_{\triangle}(s_{25};m_e^2,m_\pi^2)
\bigg)
\notag\\
&
-q_e^2\bigg(
\kappa_{12}\,I_{\triangle}(s_{12};m_e^2,m_e^2)+2\,I_{\taurus}(m_e^2)\bigg)
\notag\\
&-q_\pi^2\,
\kappa_{35}\,I_{\triangle}(s_{35};m_\pi^2,m_\pi^2)
\Bigg]\mathcal{F}^{(0)}_{i}
+ \mathcal{F}^{(1)}_{\text{fin};i} \,,
\label{eq:F_deco}
\end{align}
where $\kappa_{ij}=~2(s_{ij}-m_i^2-m_j^2)$.
Here, we use a shorthand notation for the divergent integrals in terms of two-dimensional tadpoles, $I_{\taurus}(m_i^2)$, 
and four-dimensional triangles,  
$I_{\triangle}(s_{ij};m_i^2,m_j^2)$.
Their explicit expressions are collected in Appendix~\ref{app:uv_ir_cts}.
This decomposition immediately allows us to analytically verify the $\epsilon$-poles according to the IR operator.

To make this work self-contained, ancillary files are provided~\cite{Zenodo} containing the analytic expressions up to $\mathcal{O}(\epsilon^2)$ for the bare and renormalised form factors. 
Renormalised polarised amplitudes are obtained by accounting for~\eqref{eq:FF_to_PolAmpl} (see further details in Appendix~\ref{app:anc}). 

~

Let us note that, for the purpose of evaluating the NNLO contributions, the expressions are required only up to $\mathcal{O}(\epsilon)$. However, in order to enable the complete numerical integration of all integration kernels appearing in the differential equations, solutions for the master integrals are constructed up to transcendental weight four (see Table~\ref{tab:letters}). 
The numerical evaluation of these functions is discussed in the following section.

\section{Numerical evaluation of the form factors}
\label{sec:num_eval}
With the analytic expressions for the polarised amplitudes at hand, we perform their numerical evaluation. The main source of complexity arises from computing the underlying Feynman integrals.
In this work, all Feynman integrals are expressed in terms of transcendental functions, and the differential equations they satisfy are constructed following the approach of~\cite{Caron-Huot:2014lda}. These equations are independent of $\epsilon$ and can be efficiently integrated using the numerical strategy described in~\cite{PetitRosas:2025xhm}. In particular, the canonical system of differential equations~\eqref{eq:deq_ISR} is evolved numerically, starting from a boundary point obtained with {\sc AMFlow}~\cite{Liu:2017jxz,Liu:2022chg}.

Nevertheless, two modifications to the strategy outlined in~\cite{PetitRosas:2025xhm} are required in order to evolve the system along a path that remains \textit{within the physical region}. First, all variables are evolved at the same time, and second the branch cuts are not avoided, but simply corrected to ensure a smooth numerical integration. In principle, this modifications could be avoided if one had a more complete understanding of how to enforce the Feynman $+\iu 0$ prescription when crossing to arbitrary regions. However, for processes beyond $2 \to 2$, we find this issue is not yet well understood in the literature. Therefore, we follow the strategy outlined in~\cite{Badger:2025ljy}, evolving all variables simultaneously while parametrising the integration path by a complex curve, defined as a deformation of a straight-line segment. 

While the differential equation contains rational and algebraic letters depending on the kinematic invariants $s_{ij}$ (see Sec.~\ref{sec:kin}),
the linear segments are parametrised w.r.t. $s_{12}$, $s_{35}$, the polar angles of the photon in the center-of-mass frame, and the polar and azimuthal angles of $\pi^-$ in the pion-pair rest frame, while keeping the electron and pion masses fixed. This parametrisation is then mapped onto the variables of the differential-equation system. To avoid discrete jumps induced by crossings of branch cuts associated with the square roots appearing in the system, we also employ the method of Ref.~\cite{Badger:2025ljy}. More precisely, before performing the actual evolution, we numerically determine the points at which such jumps occur and then adjust the corresponding Riemann sheets of the square roots as the path is traversed. 

It is important to stress that, although the linear path always lies within the physical region, the complex deformation must remain homotopic to it. In practice, this requires a small deformation parameter, of order $\sim 10^{-3}$. Otherwise, one may inadvertently choose a path that requires analytic continuation.
The requirement that the complex deformation be small forces some paths to pass close to singularities located on the real axis. This problem is exacerbated at low center-of-mass energies, where certain Gram determinants become small and introduce additional divergences in the letters of the system. As a result, some kinematic configurations lead to significantly longer runtimes for specific points, which can be seen in the CPU time: in a sample of $\mathcal{O}(10^4)$ phase-space points, around 0.5\% of them do not finish in less than a second. Even for phase-space points that do, the computation is more expensive than in Ref.~\cite{PetitRosas:2025xhm}, with a mean of 230ms per phase space point, since the simultaneous evolution of all variables, as opposed to an evolution carried variable-by-variable, means that the letters can no longer be precomputed. 

Furthermore, in this setup we find that numerical integration algorithms based on \texttt{Runge--Kutta} outperform \texttt{Bulirsch--Stoer}, which strongly suggests that the present system is less smooth than those studied previously, and further supports the considerations above.

A slightly tangential result, is that if one decides to use a \texttt{Runge--Kutta}-type algorithm, the branch cut crossing detection can be done on the fly, without the need to search for brackets before the integration is performed. We find this to have little impact on the runtime. Overall, improving both the CPU time and the numerical precision will require a better understanding of analytic continuation and of the structure of the physical region, which we leave for future work. 

Within this framework, we develop an in-house \texttt{C++} routine to evaluate all transcendental functions entering the form factors.

\section{Checks}
\label{sec:checks}

The analytic expressions are validated through several non-trivial checks. As a first test, the complete one-loop bare form factors are evaluated numerically using {\sc AMFlow}. In this comparison, the form factors are evaluated prior to any IBP reduction. Perfect agreement is observed for all tested phase-space points. This provides a stringent validation of the entire computational setup, with IBP reduction, construction and integration of the canonical differential equations, and the subsequent rotation to a basis of transcendental functions.

After performing UV renormalisation (see Sec.~\ref{subsection:UV-IR}), one-loop corrections to form factors (as well as polarised amplitudes) are expressed directly in the form given in Eq.~\eqref{eq:F_deco}. The IR structure is found to exactly reproduce the pole terms predicted by the action of the IR operator on tree-level contributions~\cite{Catani:2000ef}. 

To validate the finite part of the form factors, we employ {\sc GoSam}-3.0~\cite{Braun:2025afl}. Within this framework, the one-loop interference can be constructed either from the form factors,
\begin{subequations}
\begin{align}
\mathcal{M}^{(1)}
= \sum_{i,j=1}^{8} c({\boldsymbol{s},\boldsymbol{m^2}})\, 
\mathcal{F}_i^{(1)}\mathcal{F}_{j}^{(0)*}
\,,
\end{align}
or, alternatively, by summing over polarisation states,
\begin{align}
\mathcal{M}^{(1)} = 
\frac{1}{4}\sum_{h\in\vec{h}}2\,\Re\left(\mathcal{A}_{h}^{\left(1\right)}\mathcal{A}_{h}^{\left(0\right)*}\right)\,.
\end{align}
\end{subequations}
Owing to the construction of the polarised amplitudes (see Sec.~\ref{sec:kin}), we find that their analytic expressions are better suited for numerical evaluation than those based on form factors. In particular, spurious denominators involving $\mathrm{tr}_5^2$ cancel at the level of polarised amplitudes, leading to improved numerical stability.
The explicit expressions for the coefficients $c$ can be obtained by summing over spin and helicity states as well as from Eq.~\eqref{eq:FF_to_PolAmpl}, and are given in terms of rational functions of the kinematic invariants and the electron and pion masses.

This representation is fully independent of $\epsilon$, corresponding to the `t~Hooft-Veltman regularisation scheme~(HV).
This setup allows a direct comparison between the reconstructed amplitudes and the numerical evaluation provided by {\sc GoSam}. Let us remark that this comparison is performed at the level of bare amplitudes since {\sc GoSam} does not provide renormalisation for processes beyond QCD. 
The HV scheme is set in {\sc GoSam} by activating the flag {\tt regularisation\_scheme=thv} in the runcard and accounting for the normalisation difference w.r.t. our convention in Eq.~\eqref{eq:1L_int}. Numerical agreement is found for both the $\epsilon$-poles and the finite part.
In detail, approximately $\mathcal{O}(10^5)$ phase-space points are used to validate the implementation and assess the performance of the routines discussed in Sec.~\ref{sec:num_eval}.

Random momenta for the process are generated using a sequential two-body decay algorithm in the centre-of-mass frame, where the photon is first emitted isotropically, and the residual four-momentum then decays into the pion pair with angles sampled uniformly. Each phase-space point is verified against the kinematic constraints of Eqs.~(\ref{eq:kin-1}) and~(\ref{eq:kin-2}) before use.

\begin{figure}
    \centering
    \includegraphics[width=1.\linewidth]{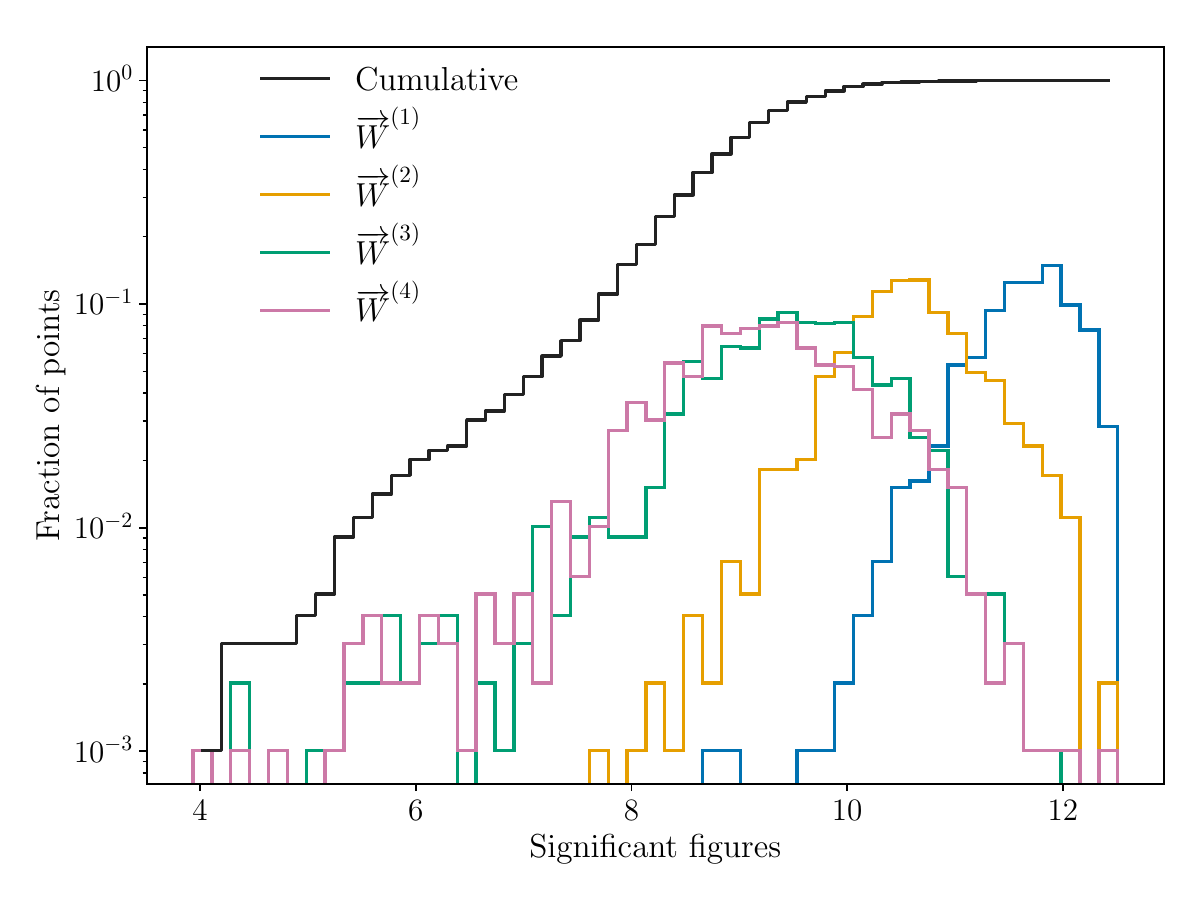}
    \caption{Significant figures reached by the integrator as compared with {\sc DiffExp} for all transcendental functions at different weights, together with the cumulative distribution over 1000 phase-space points.}
    \label{fig:prec}
\end{figure}

Furthermore, to validate the numerical evaluation of the transcendental functions at higher orders in $\epsilon$, the numerical integration of the differential equations in our implementation is compared against {\sc DiffExp}~\cite{Hidding:2020ytt}. The comparison is performed for a thousand phase space-points, with the boundary values being the same in the numerical routine and {\sc DiffExp}. We display in Fig.~\ref{fig:prec} the cumulative sum of the number of significant figures achieved by the integrator, as well as the precision of the transcendental functions at each transcendental order. In all cases, the number of significant figures is dictated by the transcendental function with the lowest precision. 

\section{Conclusions}
\label{sec:con}

We investigated the radiative return process $e^+e^-\to\pi^+\pi^-\gamma$ beyond NLO, providing essential ingredients for NNLO predictions. A complete set of four-dimensional tensor structures describing the scattering amplitudes was constructed within the 't~Hooft-Veltman regularisation scheme. A systematic procedure was developed to construct polarised amplitudes in which the dependence on Gram determinants in the denominators was eliminated, avoiding possibly numerical instabilities.

We revisited the one-loop contributions in detail and performed a series of consistency checks, including the verification of the infrared structure of the polarised amplitudes and a comparison of finite contributions against the automated implementation \textsc{GoSam}-3.0.

For the evaluation of the required Feynman integrals, significant effort has been devoted to developing a fast and reliable numerical implementation. This is achieved by numerically integrating the associated differential equations, with particular emphasis on selecting integration paths that remain entirely within the relevant physical region. For higher orders in $\epsilon$, the numerical precision of the evaluation has been validated through comparison with {\sc DiffExp}, yielding agreement at the level of 10-12 significant digits, demonstrating the suitability of this approach for integration in Monte Carlo  generators.

~

There are a number of promising directions for future work. 
We aim to compute the two-loop corrections to the radiative return process, exploiting the construction of polarised amplitudes and their organisation  into gauge-invariant subsets. 
Building on the structure of the amplitudes, we are implementing soft photon resummation within the Coherent Exclusive Exponentiation framework~\cite{Jadach:2000ir}, capturing the leading effects of the emission of multiple soft photons at both amplitude and cross-section level. 
At the same time, the numerical evaluation of transcendental functions, particularly at weights three and four, were complicated by the kinematics of the process. A key objective is therefore to have a deeper understanding of the physical region, as well as how to navigate it efficiently, avoiding getting near singularities.

\acknowledgments

We wish to thank Thomas Teubner and Graziano Venanzoni for useful discussions, and Mattia Pozzoli for collaboration on closely related projects. 
This work is supported by the Leverhulme Trust, LIP-2021-014.

\appendix

\section{UV renormalisation and IR subtraction constants}
\label{app:uv_ir_cts}

The one-loop renormalisation constants used in this calculation, up to $\mathcal{O}(\alpha)$, can be written as, 
\begin{align}
\begin{aligned}
Z_{\alpha}&=1+\frac{\alpha}{4\pi}\left(q_{F}^{2}\,\delta Z_{\alpha\left(N_{F}\right)}^{\left(1\right)}+q_{\pi}^{2}\,\delta Z_{\alpha\left(N_{S}\right)}^{\left(1\right)}\right)\,,\\Z_{\gamma}&=1+\frac{\alpha}{4\pi}\left(q_{F}^{2}\,\delta Z_{\gamma\left(N_{F}\right)}^{\left(1\right)}+q_{\pi}^{2}\,\delta Z_{\gamma\left(N_{S}\right)}^{\left(1\right)}\right)\,,\\Z_{F}&=1+\frac{\alpha}{4\pi}q_{e}^{2}\,\delta Z_{F}^{\left(1\right)}\,,\\Z_{M_{F}}&=1+\frac{\alpha}{4\pi}q_{e}^{2}\,\delta Z_{M_{F}}^{\left(1\right)}\,,\\Z_{\pi}&=1+\frac{\alpha}{4\pi}q_{\pi}^{2}\,\delta Z_{\pi}^{\left(1\right)}\,,\\Z_{M_{\pi}}&=1+\frac{\alpha}{4\pi}q_{\pi}^{2}\,\delta Z_{M_{\pi}}^{\left(1\right)}\,,
\end{aligned}
\end{align}
with,
\begin{align}
\begin{aligned}
\delta Z_{\alpha\left(N_{F}\right)}^{\left(1\right)}&=\frac{4}{3}N_{F}I_{\taurus}\left(m_{e}^{2}\right)\,,\\\delta Z_{\alpha\left(N_{S}\right)}^{\left(1\right)}&=\frac{1}{3}N_{S}I_{\taurus}\left(m_{\pi}^{2}\right)\,,\\
\delta Z_{\gamma\left(N_{F}\right)}^{\left(1\right)}&=-\delta Z_{\alpha\left(N_{F}\right)}^{\left(1\right)}\,,\\
\delta Z_{\gamma\left(N_{S}\right)}^{\left(1\right)}&=-\delta Z_{\alpha\left(N_{S}\right)}^{\left(1\right)}\,,\\
\delta Z_{F}^{\left(1\right)}&=-\frac{\left(D-1\right)}{\left(D-3\right)}I_{\taurus}\left(m_{e}^{2}\right)\,,\\
\delta Z_{M_{F}}^{\left(1\right)}&=\delta Z_{F}^{\left(1\right)}\,,\\
\delta Z_{\pi}^{\left(1\right)}&=2I_{\taurus}\left(m_{\pi}^{2}\right)\,,\\
\delta Z_{M_{\pi}}^{\left(1\right)}&=\frac{2\left(D-1\right)}{\left(D-2\right)\left(D-3\right)}I_{\taurus}\left(m_{\pi}^{2}\right)\,,
\end{aligned}
\end{align}
in terms of the tadpoles,
$I_{\taurus}\left(m_{\pi}^{2}\right)= I^{P}_{02000}$ and 
$I_{\taurus}\left(m_{e}^{2}\right)= I^{P}_{00002}$, 
with the explicit expression, 
\begin{align}
I_{\taurus}\left(m^{2}\right)=
m^{-2\epsilon}\,\frac{e^{\gamma_{\text{E}}\epsilon}\,\Gamma\left(1+\epsilon\right)}{\epsilon}
\,.
\end{align}

~

For the IR operator $\mathbf{I}$, we use a slightly modified version of the one reported in~\cite{Catani:2000ef}, with the aim of achieving a cancellation of $\epsilon$-poles at the level of Feynman integrals. The operator takes the form:
\begin{align}
\mathbf{I}_{m}&\left(\epsilon,\mu^{2};\left\{ p_{i},m_{i}\right\} \right)=
\\
&=\sum^{m}_{\substack{i,j=1
\\
i< j
}}\mathbf{T}_{i}\cdot\mathbf{T}_{j}\,
\kappa_{ij}\,
I_{\triangle}\left(s_{ij},m_{i}^{2},m_{j}^{2}\right)
-\sum_{j=1}^{m}\gamma_{j}\,I_{\taurus}\left(m_{j}^{2}\right)\,,
\notag
\end{align}
where $I_{\triangle}(s_{ij}, m_i^2, m_j^2)$ denotes the IR-divergent triangle integral with propagators $\{(\ell - q_i)^2 - m_i^2,\, \ell^2,\, (\ell + q_j)^2 - m_j^2\}$. The external momenta satisfy $q_i^2 = m_i^2$ and $q_j^2 = m_j^2$, and $s_{ij} = (q_i + q_j)^2$. The leading $\epsilon$-pole of this integral reads:  
\begin{align}
I_{\triangle}&(s_{jk}, m_j^2, m_k^2)\\
&=\frac{1}{\epsilon}\frac{1}{\kappa_{jk}v_{jk}}\left[\log\frac{1-v_{jk}}{1+v_{jk}}+2\iu\pi\Theta\left(s_{jk}\right)\right]
+\mathcal{O}\left(\epsilon^{0}\right)\,.
\notag
\end{align}
The colour factors reduce to $\mathbf{T}_i\to \pm q_i$, with $i\in\{e,\pi\}$, 
where the sign $-$ ($+$) corresponds to particles (antiparticles). 
For non-vanishing masses $m_i$ and $m_j$, the relative velocity is defined as 
\begin{align}
v_{ij} = \sqrt{1 - \frac{m_i^2 m_j^2}{(p_i \cdot p_j)^2}}\,.
\end{align}
The coefficients $\gamma_i$ are given by:
\begin{align}
\begin{aligned}
\gamma_{F} & =\frac{1}{\epsilon}q_{e}^{2}\,,
\\
\gamma_{\gamma} & =0\,,
\\
\gamma_{\pi} & =0\,.
\end{aligned}
\end{align}

\section{Ancillary files}
\label{app:anc}

The ancillary files accompanying this paper can be found in~\cite{Zenodo}.
For their organisation, all files are grouped into three folders corresponding to tree-level, initial-state, and final-state contributions, denoted by \verb"0L", \verb"1L_ISC", and \verb"1L_FSC", respectively.

The folder \verb"0L" contains:
\begin{itemize}
\item {\verb"0L_FF.m":}  tree-level form factors.
\item {\verb"0L_Pol_Ampls.m":} tree-level polarised amplitudes.
\item {\verb"FF_to_Pol_Ampls.m":} the matrices $\mathbb{H}_0=\texttt{H0}$ and $\mathbb{H}_1=\texttt{H1}$ of Eq.~\eqref{eq:FF_to_PolAmpl}, used to convert form factors into polarised amplitudes regardless of the loop order. 
\end{itemize}

The folders \verb"1L_X", with \texttt{X=ISC, FSC}, contain:
\begin{itemize}
\item {\verb"1L_X_Bare_FF.m":} bare form factors \verb"F0". 
\item {\verb"1L_X_FF.m":} renormalised form factors \verb"F".
\item {\verb"1L_X_Pol_Ampls.m":} renormalised polarised \verb"A".
\end{itemize}
All form factors and polarised amplitudes are expressed in terms of transcendental functions up to weight four. We adopt the decomposition in powers of the electron and pion charges introduced in Eq.~\eqref{eq:ISR_FSR_deco}. Explicitly, order-by-order in $\epsilon$, they are written as:
\begin{equation}
\begin{aligned}
(\mathcal{F}_{\text{X};nm}^{(L)})_i &= \sum_{k=-L}^{2L} \epsilon^k\,\texttt{F[Xnm,i][k]}\,,
\\
(\tilde{\mathcal{A}}_{\text{X};nm}^{(L)})_h &= \frac{\texttt{kinpref[h]}}{\sqrt{2}}\,\sum_{k=-L}^{2L} \epsilon^k\,\texttt{A[Xnm,h][k]}\,,
\end{aligned}
\end{equation}
with $i\in\{1,\hdots,8\}$, 
$h\in\vec{h}$ of Eq.~\eqref{eq:helicities},
$\text{X}\in\{\text{ISR, FSR}\}$,
$n,m\in\{1,2,3,4\}$, $n+m=2L+3$, and $L=0,1$. 

Since the canonical differential equation for FSC can be obtained from the ISC by crossing of kinematics, we provide it only for the ISC case. In addition, we supply the following files required for the evaluation of form factors and polarised amplitudes:
\begin{itemize}
\item{\verb"1L_ISC_DEQ.m": } 
contains the connection matrix~\eqref{eq:deq_ISR} (stored as \texttt{AtildeW}),
the letters of the kinematic alphabet (\texttt{alphabet}), 
with $\log\beta_i=\texttt{L[i]}$ and $\texttt{L[i]}=\log \texttt{b[i]}$,
the definition of square roots, \verb*|sqrts|, and the set of integrals $\overrightarrow{W}_\text{ISC} = \texttt{wfun}$.
\item{\verb"1L_X_BC.m": } 
contains the set of transcendental functions (\verb*|wfun|) and the boundary conditions \verb"bc[ptn]" at three phase-space points \verb"x[ptn]", with \texttt{n=1,2,3}.
\item{\verb"1L_X_MIs.m": } 
contains the set of master integrals (\verb*|MIs|), obtained from the analysis of their leading and Landau singularities, together with the definition of square roots \verb*|sqrts|, 
and the rotation to $\overrightarrow{W}_\text{X}$ (stored in \texttt{UTtoW}).
\end{itemize}

\bibliographystyle{JHEP}
\bibliography{refs}

\end{document}